\documentclass[english,pccp,preprint,superscriptaddress]{revtex4-1} 
\usepackage{epsfig}
\usepackage{amsmath} 
\usepackage{graphicx} 
\usepackage{longtable} 
\usepackage{hyperref} 
\usepackage{amssymb} 
\usepackage{dcolumn} 
\usepackage{mhchem} 
\usepackage{cleveref} 
\usepackage{subfigure} 
\usepackage{array} 
\usepackage{multirow} 
\usepackage{tabularx} 
\usepackage{cleveref} 
\makeatletter
\renewcommand*{\@fnsymbol}[1]{\ensuremath{\ifcase#1\or \dagger\or *\or \ddagger\or 
   \mathsection\or \mathparagraph\or \|\or **\or \dagger\dagger
   \or \ddagger\ddagger \else\@ctrerr\fi}}
\makeatother

\begin{document} 

\title{Electronic Properties of Cyclacenes from TAO-DFT} 

\author{Chun-Shian Wu} 
\thanks{These authors contributed equally to this work.} 
\affiliation{Department of Physics, National Taiwan University, Taipei 10617, Taiwan} 
\affiliation{Department of Chemistry, National Taiwan University, Taipei 10617, Taiwan} 

\author{Pei-Yin Lee} 
\thanks{These authors contributed equally to this work.} 
\affiliation{Department of Physics, National Taiwan University, Taipei 10617, Taiwan} 

\author{Jeng-Da Chai} 
\email[Author to whom correspondence should be addressed. Electronic mail: ]{jdchai@phys.ntu.edu.tw} 
\affiliation{Department of Physics, National Taiwan University, Taipei 10617, Taiwan} 
\affiliation{Center for Theoretical Sciences and Center for Quantum Science and Engineering, National Taiwan University, Taipei 10617, Taiwan} 

\date{\today} 

\begin{abstract} 

Owing to the presence of strong static correlation effects, accurate prediction of the electronic properties (e.g., the singlet-triplet energy gaps, vertical ionization potentials, vertical 
electron affinities, fundamental gaps, symmetrized von Neumann entropy, active orbital occupation numbers, and real-space representation of active orbitals) of cyclacenes with $n$ 
fused benzene rings ($n$ = 4--100) has posed a great challenge to traditional electronic structure methods. To meet the challenge, we study these properties using our newly developed 
thermally-assisted-occupation density functional theory (TAO-DFT), a very efficient method for the study of large systems with strong static correlation effects. Besides, to examine the 
role of cyclic topology, the electronic properties of cyclacenes are also compared with those of acenes. Similar to acenes, the ground states of cyclacenes are singlets for all the cases 
studied. In contrast to acenes, the electronic properties of cyclacenes, however, exhibit oscillatory behavior (for $n \le 30$) in the approach to the corresponding properties of acenes with 
increasing number of benzene rings. On the basis of the calculated orbitals and their occupation numbers, the larger cyclacenes are shown to exhibit increasing polyradical character in 
their ground states, with the active orbitals being mainly localized at the peripheral carbon atoms. 

\end{abstract} 

\maketitle 

\section*{Introduction} 

Carbon nanotubes (CNTs) are promising nanomaterials, which have been extensively studied by many researchers \cite{cnt1,cnt2,cnt3,cnt4,cnt5,cnt6,cnt7,cnt8,cnt9,cnt10,25,27,28,23,16}. 
Due to different combinations of structural variation, CNTs can exhibit a wide range of electronic and optical properties, which can be of great use in the design of novel techniques \cite{25}. 
CNTs are also polyfunctional macromolecules, where specific reactions can occur at various sites with different efficiencies \cite{28}. There are three major types of CNTs: armchair CNTs, 
chiral CNTs, and zigzag CNTs, which are distinguished by the geometrical vector ($n$,$m$), with $n$ and $m$ being integers. CNTs can behave as either metals or semiconductors 
depending on their chiral angles, diameters, and lengths. Therefore, a further investigation of how these factors affect the properties of CNTs is essential for the comprehensive 
understanding of these materials \cite{25,23}. 

\begin{figure} 
\includegraphics[scale=0.7]{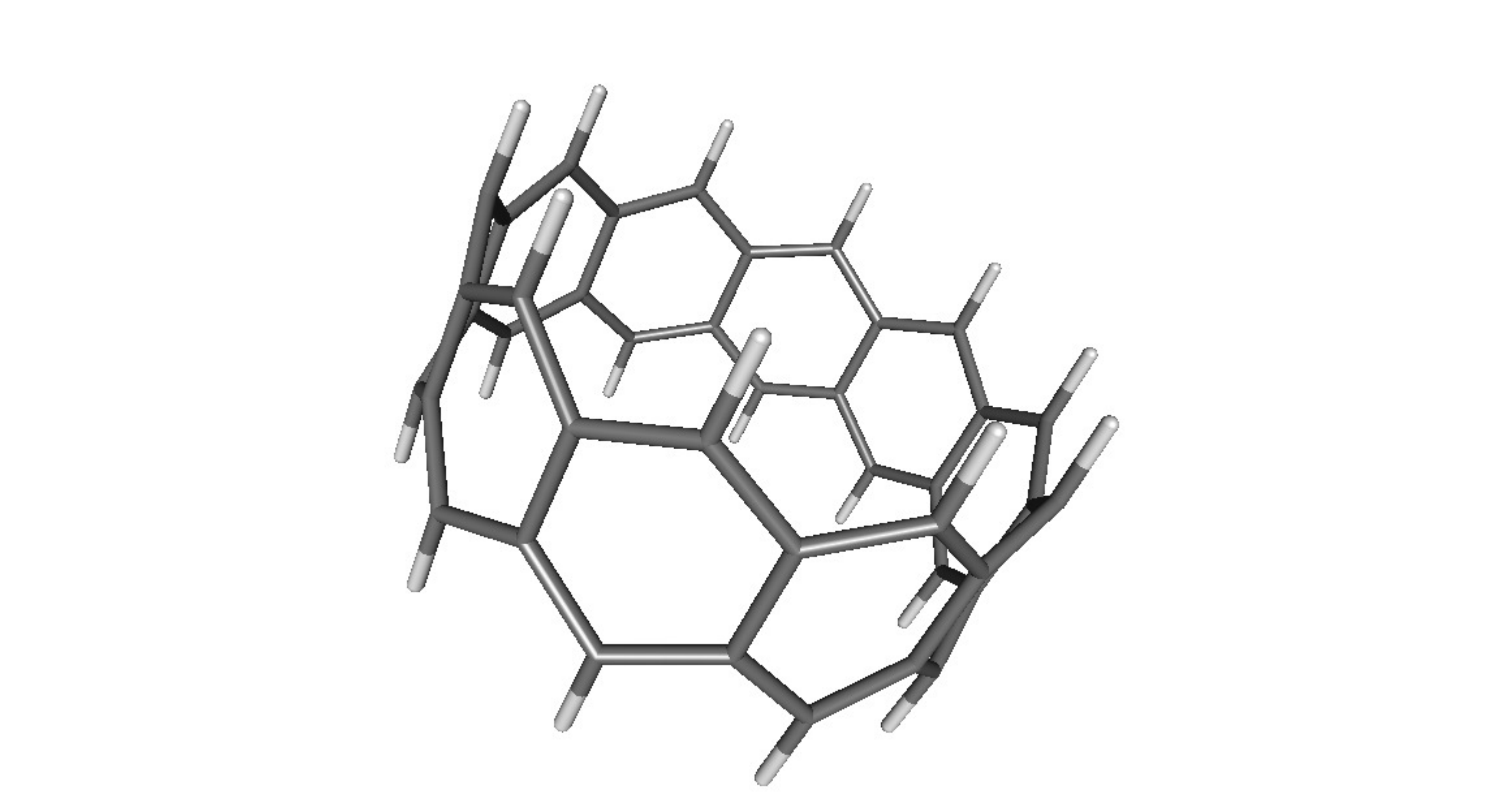} 
\caption{\label{fig:tube-cyclic10-geometry} 
Structure of 10-cyclacene, consisting of 10 fused benzene rings forming a closed loop.} 
\end{figure}

In particular, it is useful to study the basic repeating units of CNTs, which still need further fundamental research exploration \cite{16}. The targeting units of the present study, a series of 
$n$-cyclacenes, consisting of $n$ fused benzene rings forming a closed loop (see \Cref{fig:tube-cyclic10-geometry}), are the shortest ($n$,0) zigzag CNTs with hydrogen passivation, 
which have attracted considerable interest in the research community due to their fascinating electronic properties \cite{23,16,7,19,31,1,6,11,24,17,10,9,C2015,new1}. As $n$-cyclacenes 
belong to the category of cata-condensed aromatics (i.e., molecules that have no carbon atoms belonging to more than two rings), each carbon atom is on the periphery of the conjugated 
system \cite{7}. Before $n$-cyclacenes are intensively connected to zigzag CNTs, they have been studied mainly due to the research curiosity in highly conjugated cyclic systems. The 
studies of $n$-cyclacenes can also be important for atomic-level structural control in the synthesis of CNTs. In addition, bottom-up approaches to the synthesis of CNTs not only provide a 
fundamental understanding of the relationship between the design of CNTs and their electronic properties, but also greatly lower the synthetic temperatures \cite{25}. While zigzag CNTs 
may be synthesized from cycloarylenes by devising the cutout positions of CNTs \cite{27}, it remains important to systematically investigate the properties of $n$-cyclacenes, which can be 
useful for exploring the possible utility of their cylindrical cavities in host-guest chemistry \cite{31}. 

The structure of $n$-cyclacene has two types of components: an arenoid belt (composed of fused benzene rings) and two peripheral circuits (the top and bottom peripheral circuits) \cite{1}. 
The peripheral circuits are of two types: $4k$ and $4k+2$ (where $k$ is an integer), depending on the number of benzene rings in $n$-cyclacene. In previous studies, it has been shown 
that $n$-cyclacene with even-number benzene rings ($4k$ type) is more stable than that with odd-number benzene rings ($4k+2$ type) \cite{1,7,6,new1}. Therefore, the nature of 
peripheral circuits (i.e., the cryptoannulenic effect) is expected to be responsible for the properties of $n$-cyclacene. Besides, the structure of $n$-cyclacene can also be regarded as two 
fused trannulenes (i.e., circular, all-trans cyclic polyene ribbons) \cite{31,11}. From the bond length analysis of $n$-cyclacene, there is bond length alternation in the benzene ring, and the 
aromaticity is reduced due to the structural strain, which can hence be responsible for the properties of $n$-cyclacene. 

Even though there has been a keen interest in $n$-cyclacenes, the studies of their electronic properties are scarce. While $n$-cyclacene may be synthesized via an intramolecular 
cyclization of $n$-acene (a chain-like molecule with $n$ linearly fused benzene rings, e.g., see Figure 1 of Ref.\ \cite{1}), the synthetic procedure has been very challenging, and has not 
succeeded in producing pure $n$-cyclacene \cite{17,1,10}, possibly due to its highly strained structure and highly reactive nature \cite{16,17}. As the stabilities of annulated polycyclic 
saturated hydrocarbons decrease rapidly with the number of fused benzene rings \cite{19}, the synthesis of larger $n$-cyclacenes should be even more difficult. 

To date, the reported properties of $n$-cyclacenes are based on theoretical calculations. Nevertheless, accurate prediction of the electronic properties of larger $n$-cyclacenes has been 
very challenging for traditional electronic structure methods, due to the presence of strong static correlation effects \cite{9}. Kohn-Sham density functional theory (KS-DFT) \cite{ks2} with 
conventional (i.e., semilocal \cite{kslda1,kslda2,PBE,M06L}, hybrid \cite{hybrid,B3LYP,LCHirao,wB97X,wB97X-D,wM05-D,LC-D3,LCAC}, and 
double-hybrid \cite{B2PLYP,wB97X-2,PBE0-2,SCAN0-2}) exchange-correlation (XC) density functionals can yield unreliable results for systems with strong static correlation 
effects \cite{Cohen2012}. High-level {\it ab initio} multi-reference 
methods \cite{9,CASPT2,Acene-DMRG,2-RDM,GNRs-DMRG,GNRs-PHF,GNRs-MRAQCC,multi-reference,multi-reference2} are typically required to accurately predict the properties of 
larger $n$-cyclacenes. However, as the number of electrons in $n$-cyclacene quickly increases with increasing $n$, there have been very few studies on the properties of larger 
$n$-cyclacenes using multi-reference methods, due to their prohibitively high cost. 

To circumvent the formidable computational expense of high-level {\it ab initio} multi-reference methods, we have recently developed thermally-assisted-occupation density functional theory 
(TAO-DFT) \cite{tao1,tao2}, a very efficient electronic structure method for studying the properties of large ground-state systems (e.g., containing up to a few thousand electrons) with strong 
static correlation effects \cite{z,NK,HSM}. 
In contrast to KS-DFT, TAO-DFT is a density functional theory with fractional orbital occupations, wherein strong static correlation is explicitly described by the entropy contribution (see 
Eq.\ (26) of Ref.\ \cite{tao1}), a function of the fictitious temperature and orbital occupation numbers. Note that the entropy contribution is completely missing in KS-DFT. Recently, we have 
studied the electronic properties of zigzag graphene nanoribbons (ZGNRs) using TAO-DFT \cite{z}. The ground states of ZGNRs are found to be singlets for all the widths and lengths 
studied. The longer ZGNRs should possess increasing polyradical character in their ground states, with the active orbitals being mainly localized at the zigzag edges. Our results are in 
good agreement with the available experimental and highly accurate {\it ab initio} data. Besides, on the basis of our TAO-DFT calculations, the active orbital occupation numbers for the 
ground states of ZGNRs should exhibit a curve crossing behavior in the approach to unity (singly occupied) with increasing ribbon length. Very recently, the curve crossing behavior has 
been confirmed by highly accurate {\it ab initio} multi-reference methods \cite{multi-reference2}! 

TAO-DFT has similar computational cost as KS-DFT for single-point energy and analytical nuclear gradient calculations, and reduces to KS-DFT in the absence of strong static correlation 
effects. Besides, existing XC density functionals in KS-DFT may also be adopted in TAO-DFT. Relative to high-level {\it ab initio} multi-reference methods, TAO-DFT is computationally 
efficient, and hence very powerful for the study of large polyradical systems. In addition, the orbital occupation numbers from TAO-DFT, which are intended to simulate the natural orbital 
occupation numbers (NOONs) [i.e., the eigenvalues of one-electron reduced density matrix] \cite{noon}, can be very useful for assessing the possible polyradical character of systems. 
Recent studies have demonstrated that the orbital occupation numbers from TAO-DFT are qualitatively similar to the NOONs from high-level {\it ab initio} multi-reference methods, giving 
promise for applying TAO-DFT to large polyradical systems \cite{tao1,z,NK,multi-reference2}. 

Due to its computational efficiency and reasonable accuracy for large systems with strong static correlation effects, in this work, TAO-DFT is adopted to study the electronic properties of 
$n$-cyclacenes ($n$ = 4--100). As $n$-cyclacenes have not been successfully synthesized, no experimental data are currently available for comparison. Therefore, our results are 
compared with the available high-level {\it ab initio} data as well as those obtained from various XC density functionals in KS-DFT. In addition, as $n$-cyclacene can be considered as an 
interconnection of $n$-acene, the electronic properties of $n$-cyclacene are also compared with those of $n$-acene to assess the role of cyclic topology.

\section*{Computational Details} 

All calculations are performed with a development version of \textsf{Q-Chem 4.0} \cite{qchem}, using the 6-31G(d) basis set with the fine grid EML(75,302), consisting of 75 Euler-Maclaurin 
radial grid points and 302 Lebedev angular grid points. Results are calculated using KS-LDA (i.e., KS-DFT with the LDA XC density functional \cite{kslda1,kslda2}) and TAO-LDA (i.e., 
TAO-DFT with the LDA XC density functional and the LDA $\theta$-dependent density functional $E_{\theta}^{\text {LDA}}$ (see Eq.\ (41) of Ref.\ \cite{tao1}) with the fictitious temperature 
$\theta$ = 7 mhartree (as defined in Ref.\ \cite{tao1}). Note that KS-LDA is simply TAO-LDA with $\theta$ = 0, and hence it is important to assess the performance of KS-LDA here to 
assess the significance of TAO-LDA. 

The ground state of $n$-cyclacene/$n$-acene ($n$ = 4--100) is obtained by performing spin-unrestricted KS-LDA and TAO-LDA calculations for the lowest singlet and triplet energies of 
$n$-cyclacene/$n$-acene on the respective geometries that were fully optimized at the same level of theory. The singlet-triplet energy (ST) gap of $n$-cyclacene/$n$-acene is calculated 
as $(E_{\text{T}} - E_{\text{S}})$, the energy difference between the lowest triplet (T) and singlet (S) states of $n$-cyclacene/$n$-acene.

\section*{Results and Discussion} 

\subsection*{Singlet-Triplet Energy Gap} 

\begin{figure} 
\includegraphics[scale=0.8]{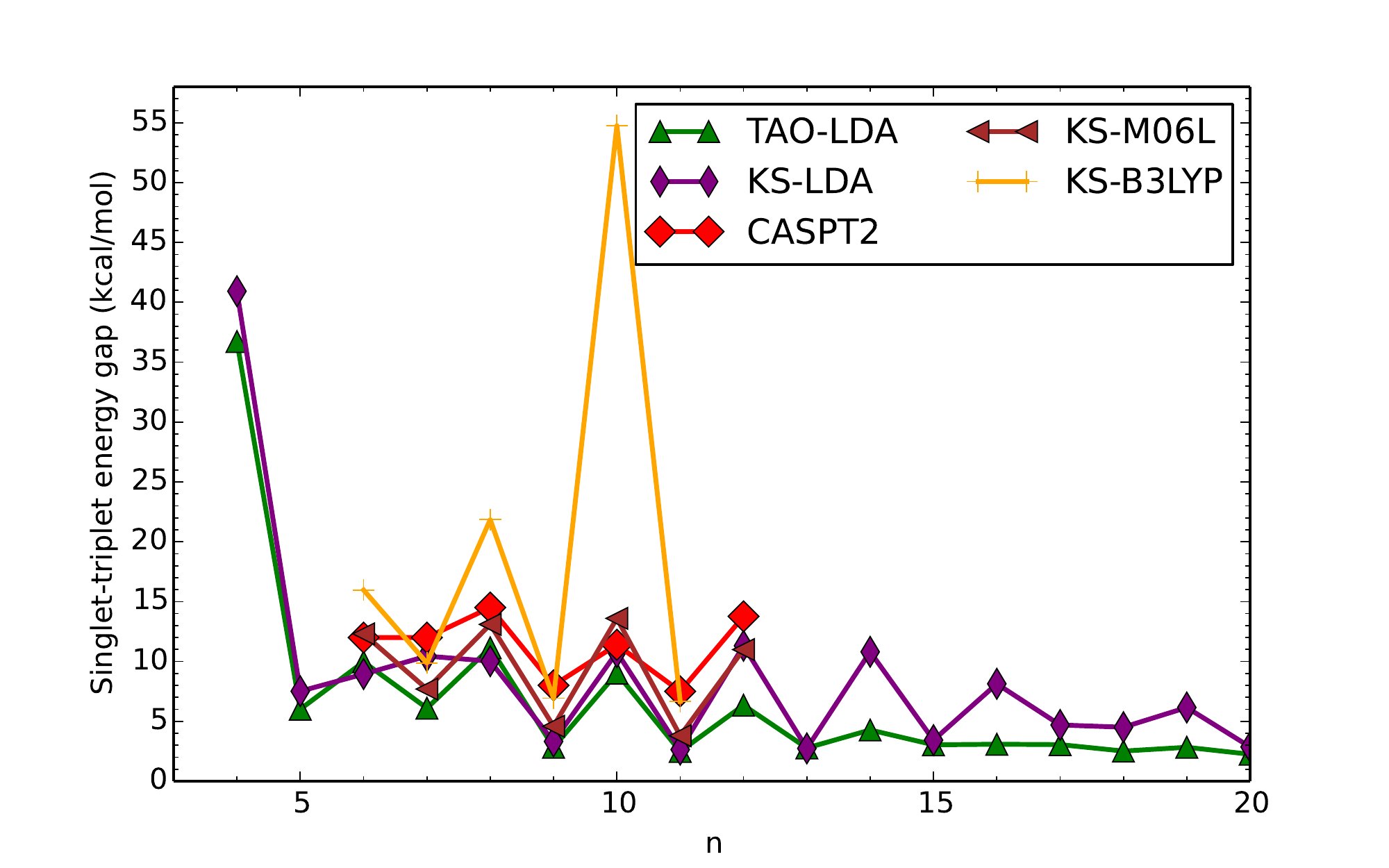} 
\caption{\label{fig:stgap} 
Singlet-triplet energy gap of $n$-cyclacene as a function of the number of benzene rings, calculated using TAO-LDA and KS-LDA. 
For comparison, the CASPT2, KS-M06L, and KS-B3LYP data are taken from the literature \cite{9}.} 
\end{figure}

\Cref{fig:stgap} shows the ST gap of $n$-cyclacene as a function of the number of benzene rings, calculated using spin-unrestricted KS-LDA and TAO-LDA. The results are compared with 
the available data \cite{9}, calculated using the complete-active-space second-order perturbation theory (CASPT2) \cite{CASPT2} (a high-level {\it ab initio} multi-reference method) as well 
as the M06L functional \cite{M06L} (a popular semilocal XC density functional) and the B3LYP functional \cite{hybrid,B3LYP} (a popular hybrid XC density functional) in KS-DFT. 

As can be seen, the anticipated even-odd oscillations in the ST gaps may be attributed to the cryptoannulenic effects of $n$-cyclacenes \cite{1,7,6,new1}. However, the amplitudes of the 
even-odd oscillations are considerably larger for KS-DFT with the XC functionals, which are closely related to the degree of spin contamination (as discussed in Ref.\ \cite{9}). In general, 
the larger fraction of Hartree-Fock (HF) exchange adopted in the XC functional in KS-DFT, the higher the degree of spin contamination for systems with strong static correlation effects. 
For example, the ST gap obtained with KS-B3LYP is unexpectedly large at $n$ = 10, due to the high degree of spin contamination \cite{9}. 

On the other hand, as commented in Ref.\ \cite{9}, the ST gaps obtained with CASPT2 are rather sensitive to the choice of active space. Since the complete $\pi$-valence space was not 
selected as the active space (due to the prohibitively high cost), the CASPT2 results here should be taken with caution. Recent studies have shown that a sufficiently large active space 
should be adopted in high-level {\it ab initio} multi-reference calculations \cite{Acene-DMRG,GNRs-DMRG,multi-reference2} for accurate prediction of the electronic properties of systems 
with strong static correlation effects, which can, however, be prohibitively expensive for large systems. Note that the ST gap obtained with CASPT2 unexpectedly increases at $n$ = 12, 
possibly due to the insufficiently large active space adopted in the calculations \cite{9}. 

\begin{figure} 
\includegraphics[scale=0.8]{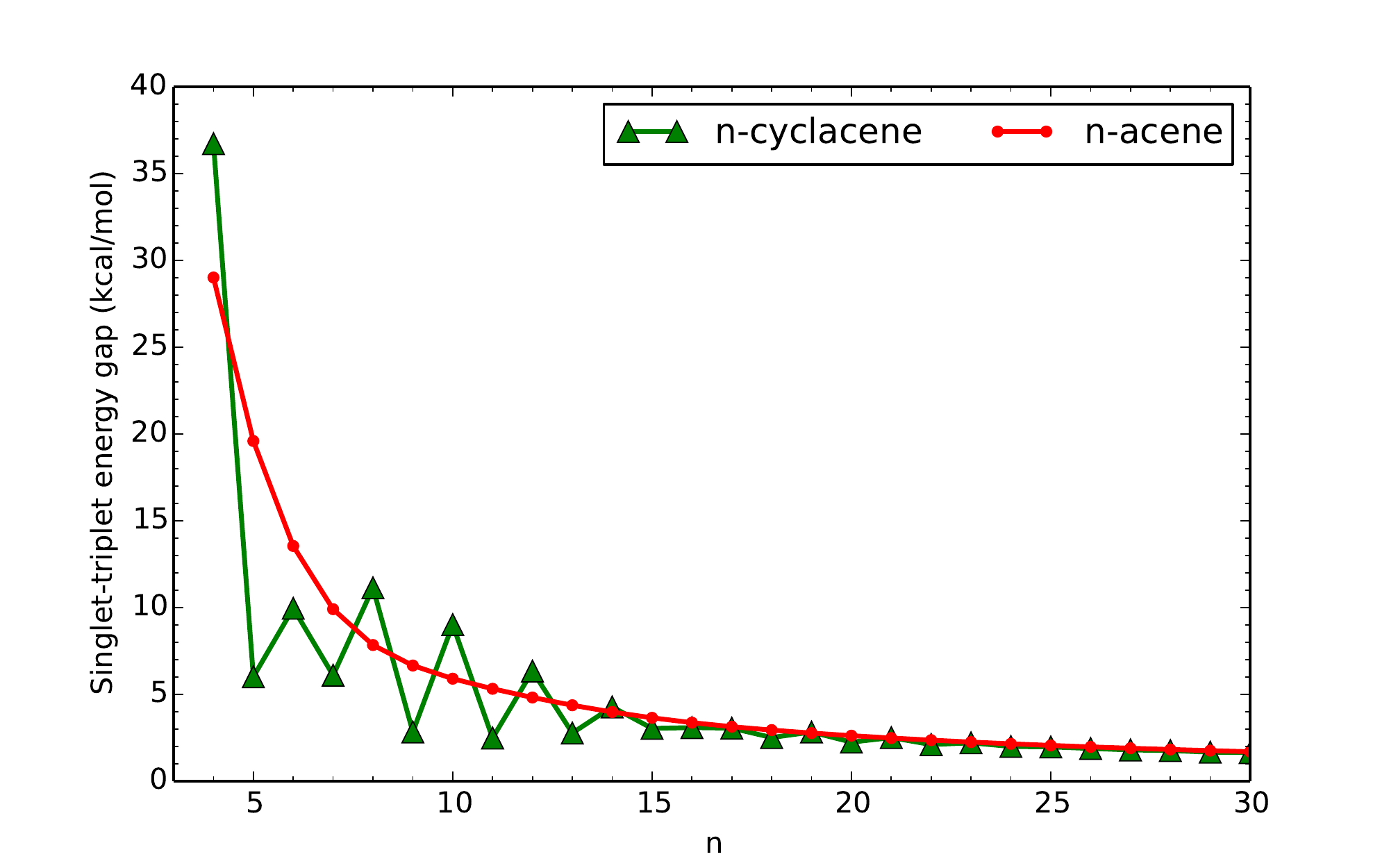} 
\caption{\label{fig:stgapcycace1} 
Singlet-triplet energy gap of $n$-cyclacene/$n$-acene as a function of the number of benzene rings, calculated using TAO-LDA.} 
\end{figure}

\begin{figure} 
\includegraphics[scale=0.8]{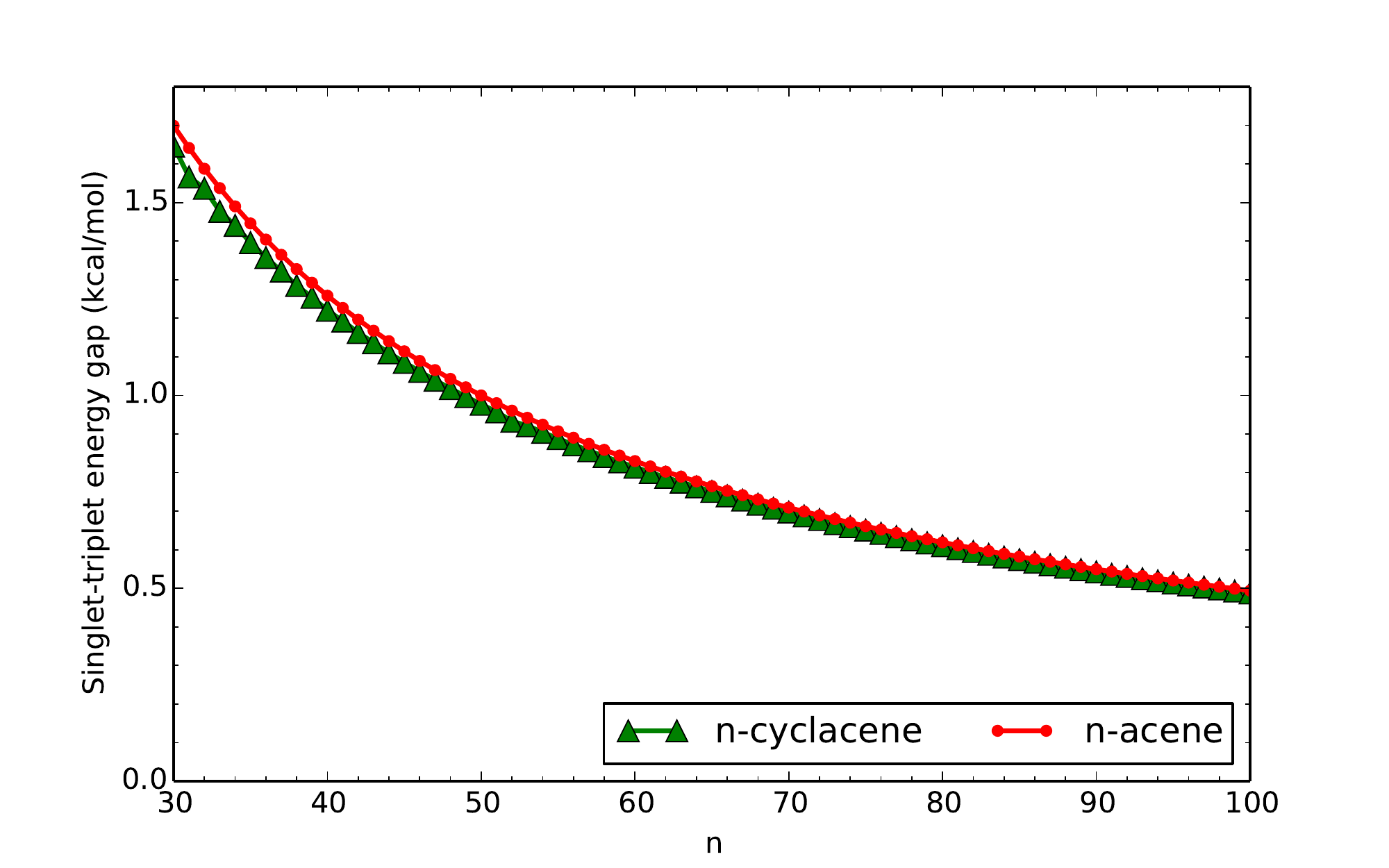} 
\caption{\label{fig:stgapcycace2} 
Same as \Cref{fig:stgapcycace1}, but for the larger $n$-cyclacene/$n$-acene.} 
\end{figure}

To assess the role of cyclic topology, \Cref{fig:stgapcycace1,fig:stgapcycace2} show the ST gap of $n$-cyclacene/$n$-acene as a function of the number of benzene rings, calculated with 
spin-unrestricted TAO-LDA. Similar to $n$-acenes, the ground states of $n$-cyclacenes remain singlets for all the cases investigated. In contrast to $n$-acene, the ST gap of 
$n$-cyclacene, however, displays oscillatory behavior for small $n$, and the oscillation vanishes gradually with increasing $n$. For small $n$, $n$-cyclacene with even-number benzene 
rings exhibits a larger ST gap (i.e., greater stability) than that with odd-number benzene rings. For sufficiently large $n$ ($n > 30$), the ST gap of $n$-cyclacene converges monotonically 
from below to the ST gap of $n$-acene (which monotonically decreases with increasing $n$). At the level of TAO-LDA, the ST gaps of the largest $n$-cyclacene and $n$-acene studied 
(i.e., $n$ = 100) are essentially the same (0.49 kcal/mol). On the basis of the ST gaps obtained with TAO-LDA, the cryptoannulenic effect and structural strain of $n$-cyclacene are more 
important for the smaller $n$, and less important for the larger $n$. 

Due to the symmetry constraint, the spin-restricted and spin-unrestricted energies for the lowest singlet state of $n$-cyclacene/$n$-acene, calculated using the exact theory, should be 
identical \cite{tao1,tao2,z,GNRs-PHF}. Recent studies have shown that KS-DFT with conventional XC density functionals cannot satisfy this condition for the larger 
$n$-cyclacene/$n$-acene, due to the aforementioned spin contamination \cite{9,Acene-DMRG,GNRs-DMRG,GNRs-PHF,tao1,tao2,z}. To assess the possible symmetry-breaking effects, 
spin-restricted TAO-LDA calculations are also performed for the lowest singlet energies on the respective optimized geometries. Within the numerical accuracy of our calculations, the 
spin-restricted and spin-unrestricted TAO-LDA energies for the lowest singlet state of $n$-cyclacene/$n$-acene are essentially the same (i.e., essentially no unphysical symmetry-breaking 
effects occur in our spin-unrestricted TAO-LDA calculations).

\subsection*{Vertical Ionization Potential, Vertical Electron Affinity, and Fundamental Gap} 

At the lowest singlet state (i.e., the ground-state) geometry of $n$-cyclacene/$n$-acene (containing $N$ electrons), TAO-LDA is adopted to calculate the vertical ionization potential 
$\text{IP}_{v}={E}_{N-1}-{E}_{N}$, vertical electron affinity $\text{EA}_{v}={E}_{N}-{E}_{N+1}$, and fundamental gap $E_{g}=\text{IP}_{v}-\text{EA}_{v}={E}_{N+1}+{E}_{N-1}-2{E}_{N}$ 
using multiple energy-difference methods, with ${E}_{N}$ being the total energy of the $N$-electron system. 

\begin{figure} 
\includegraphics[scale=0.8]{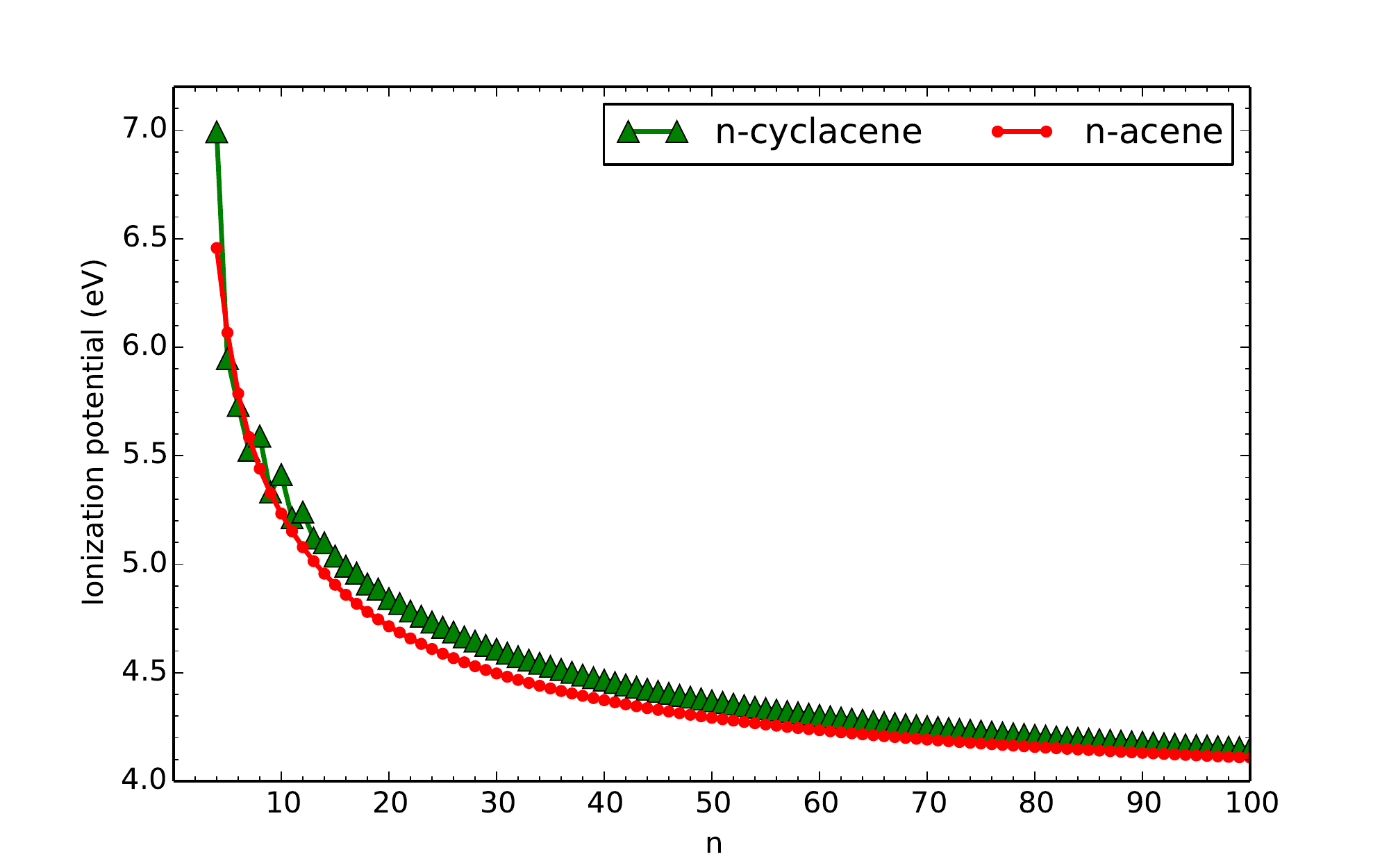} 
\caption{\label{fig:ip} 
Vertical ionization potential for the lowest singlet state of $n$-cyclacene/$n$-acene as a function of the number of benzene rings, calculated using TAO-LDA.} 
\end{figure}

\begin{figure} 
\includegraphics[scale=0.8]{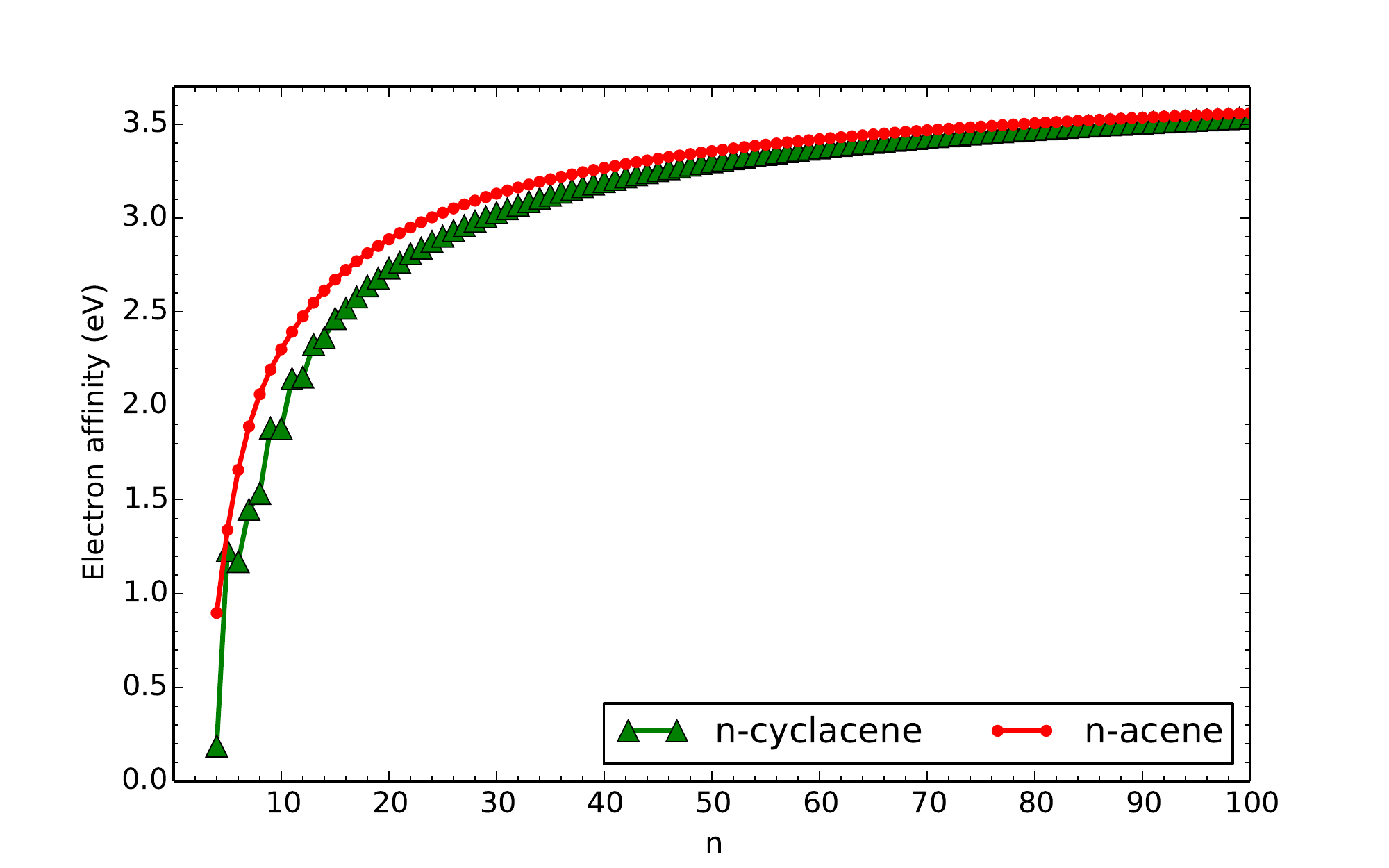} 
\caption{\label{fig:ea} 
Vertical electron affinity for the lowest singlet state of $n$-cyclacene/$n$-acene as a function of the number of benzene rings, calculated using TAO-LDA.} 
\end{figure}

\begin{figure} 
\includegraphics[scale=0.8]{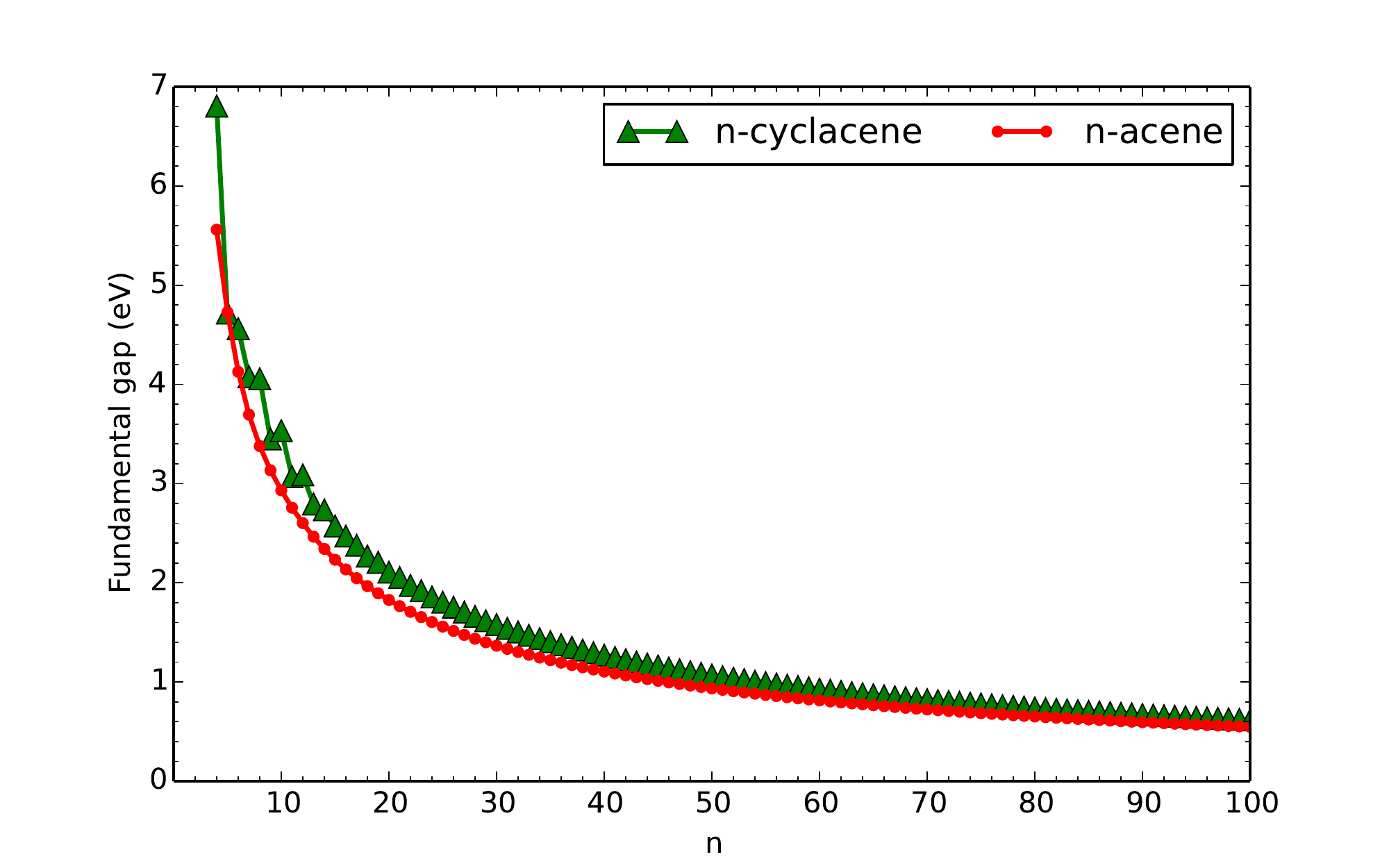} 
\caption{\label{fig:fg} 
Fundamental gap for the lowest singlet state of $n$-cyclacene/$n$-acene as a function of the number of benzene rings, calculated using TAO-LDA.} 
\end{figure}

With increasing number of benzene rings in $n$-cyclacene, $\text{IP}_{v}$ oscillatorily decreases (see \Cref{fig:ip}), $\text{EA}_{v}$ oscillatorily increases (see \Cref{fig:ea}), and hence 
$E_{g}$ oscillatorily decreases (see \Cref{fig:fg}). However, these oscillations are damped and eventually disappear with increasing $n$. For sufficiently large $n$ ($n > 30$), the 
$\text{IP}_{v}$ and $E_{g}$ values of $n$-cyclacene converge monotonically from above to those of $n$-acene (which monotonically decrease with increasing $n$), while the 
$\text{EA}_{v}$ value of $n$-cyclacene converges monotonically from below to that of $n$-acene (which monotonically increases with increasing $n$). Note also that the $E_{g}$ value 
of $n$-cyclacene ($n$ = 13--54) is within the most interesting range (1 to 3 eV), giving promise for applications of $n$-cyclacenes in nanophotonics.

\subsection*{Symmetrized von Neumann Entropy} 

To investigate the possible polyradical character of $n$-cyclacene/$n$-acene, we calculate the symmetrized von Neumann entropy (e.g., see Eq.\ (9) of Ref.\ \cite{GNRs-PHF}) 
\begin{equation}\label{eq1} 
S_{\text{vN}} = -\frac{1}{2} \sum_{i=1}^{\infty} \bigg\lbrace f_{i}\ \text{ln}(f_{i}) + (1-f_{i})\ \text{ln}(1-f_{i}) \bigg\rbrace, 
\end{equation} 
for the lowest singlet state of $n$-cyclacene/$n$-acene as a function of the number of benzene rings, using TAO-LDA. Here, $f_{i}$ the occupation number of the $i^{\text{th}}$ orbital 
obtained with TAO-LDA, which ranges from 0 to 1, is approximately the same as the occupation number of the $i^{\text{th}}$ natural orbital \cite{tao1,tao2,z,NK,HSM,multi-reference2}. 
For a system without strong static correlation ($\{f_{i}\}$ are close to either 0 or 1), $S_{\text{vN}}$ provides insignificant contributions, while for a system with strong static correlation 
($\{f_{i}\}$ are fractional for active orbitals and are close to either 0 or 1 for others), $S_{\text{vN}}$ increases with the number of active orbitals. 

\begin{figure} 
\includegraphics[scale=0.8]{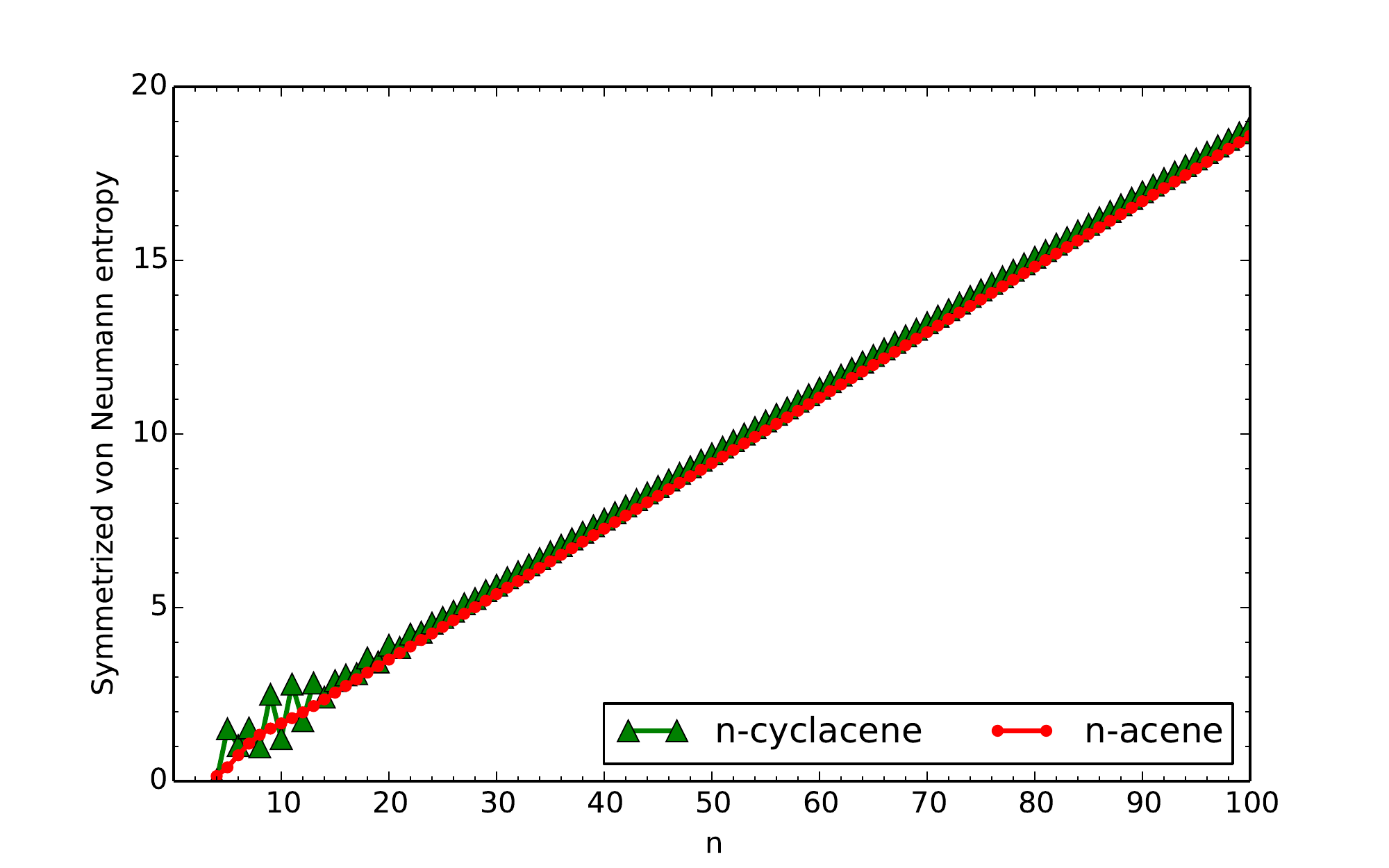} 
\caption{\label{fig:s} 
Symmetrized von Neumann entropy for the lowest singlet state of $n$-cyclacene/$n$-acene as a function of the number of benzene rings, calculated using TAO-LDA.} 
\end{figure}

As shown in \Cref{fig:s}, the $S_{\text{vN}}$ value of $n$-cyclacene oscillatorily increases with increasing number of benzene rings. Nonetheless, the oscillation is damped and eventually 
disappears with the increase of $n$. For sufficiently large $n$ ($n > 30$), the $S_{\text{vN}}$ value of $n$-cyclacene converges monotonically from above to that of $n$-acene (which 
monotonically increases with increasing $n$). Therefore, similar to $n$-acenes \cite{tao1,tao2,z,NK,HSM,Acene-DMRG,GNRs-DMRG,GNRs-PHF,GNRs-MRAQCC,multi-reference2}, the 
larger $n$-cyclacenes should possess increasing polyradical character.

\subsection*{Active Orbital Occupation Numbers} 

To illustrate the causes of the increase of $S_{\text{vN}}$ with $n$, we plot the active orbital occupation numbers for the lowest singlet state of $n$-cyclacene as a function of the number 
of benzene rings, calculated using TAO-LDA. Here, the highest occupied molecular orbital (HOMO) is the ${(N/2)}^{\text{th}}$ orbital, and the lowest unoccupied molecular orbital (LUMO) 
is the ${(N/2 + 1)}^{\text{th}}$ orbital, where $N$ is the number of electrons in $n$-cyclacene. For brevity, HOMO, HOMO$-$1, ..., and HOMO$-$15, are denoted as H, H$-$1, ..., and 
H$-$15, respectively, while LUMO, LUMO+1, ..., and LUMO+15, are denoted as L, L+1, ..., and L+15, respectively. 

\begin{figure} 
\includegraphics[scale=0.45]{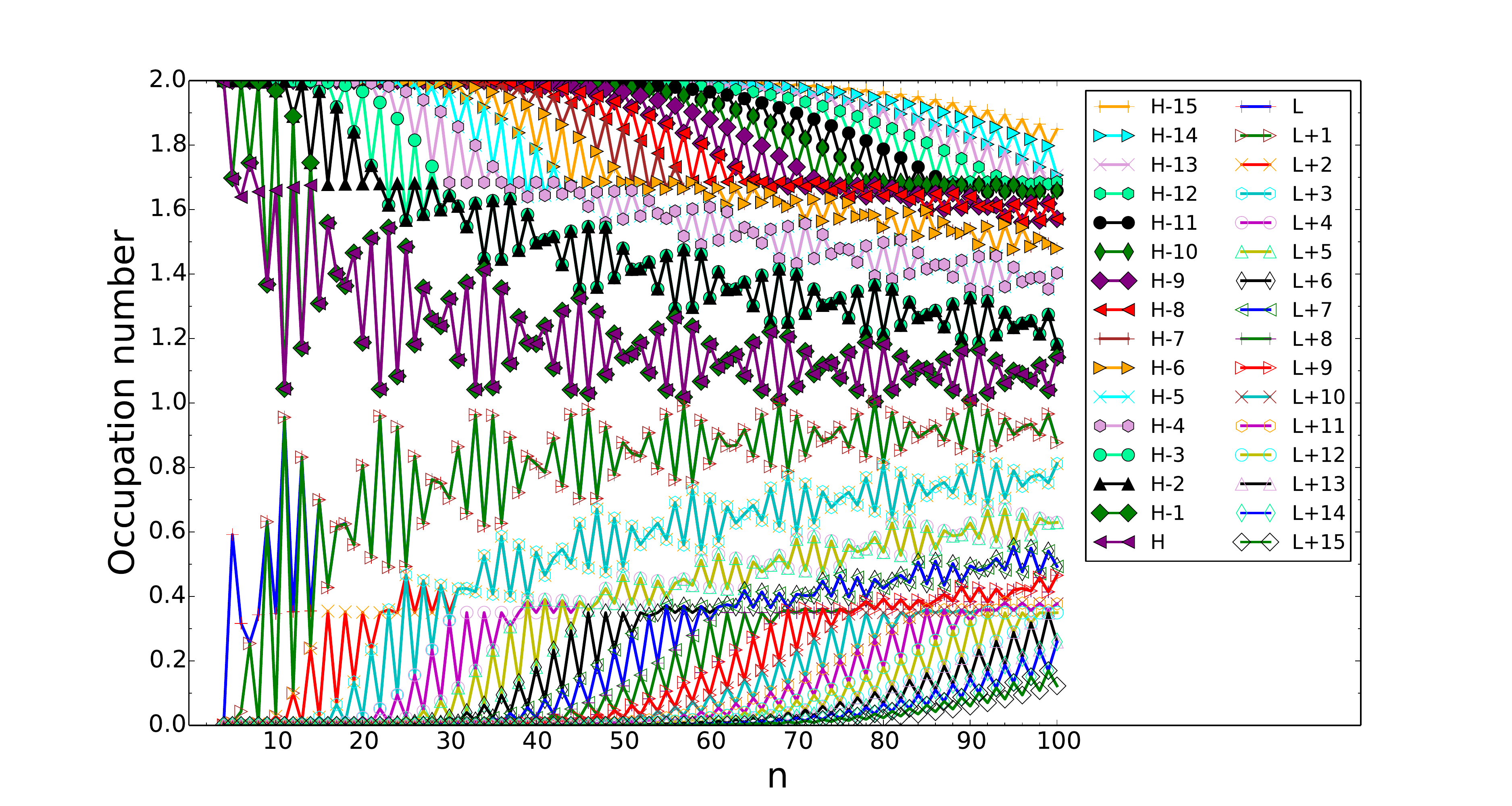} 
\caption{\label{fig:occupation} 
Active orbital occupation numbers (HOMO$-$15, ..., HOMO$-$1, HOMO, LUMO, LUMO+1, ..., and LUMO+15) for the lowest singlet state of $n$-cyclacene 
as a function of the number of benzene rings, calculated using TAO-LDA. For brevity, HOMO is denoted as H, LUMO is denoted as L, and so on.} 
\end{figure}

As presented in \Cref{fig:occupation}, the number of fractionally occupied orbitals increases with increasing cyclacene size, clearly indicating that the polyradical character of 
$n$-cyclacene indeed increases with the cyclacene size. Similar to the previously discussed properties, the active orbital occupation numbers of $n$-cyclacene also exhibit oscillatory 
behavior, showing wave-packet oscillations.

\subsection*{Real-Space Representation of Active Orbitals} 

\begin{figure} 
\includegraphics[scale=0.7]{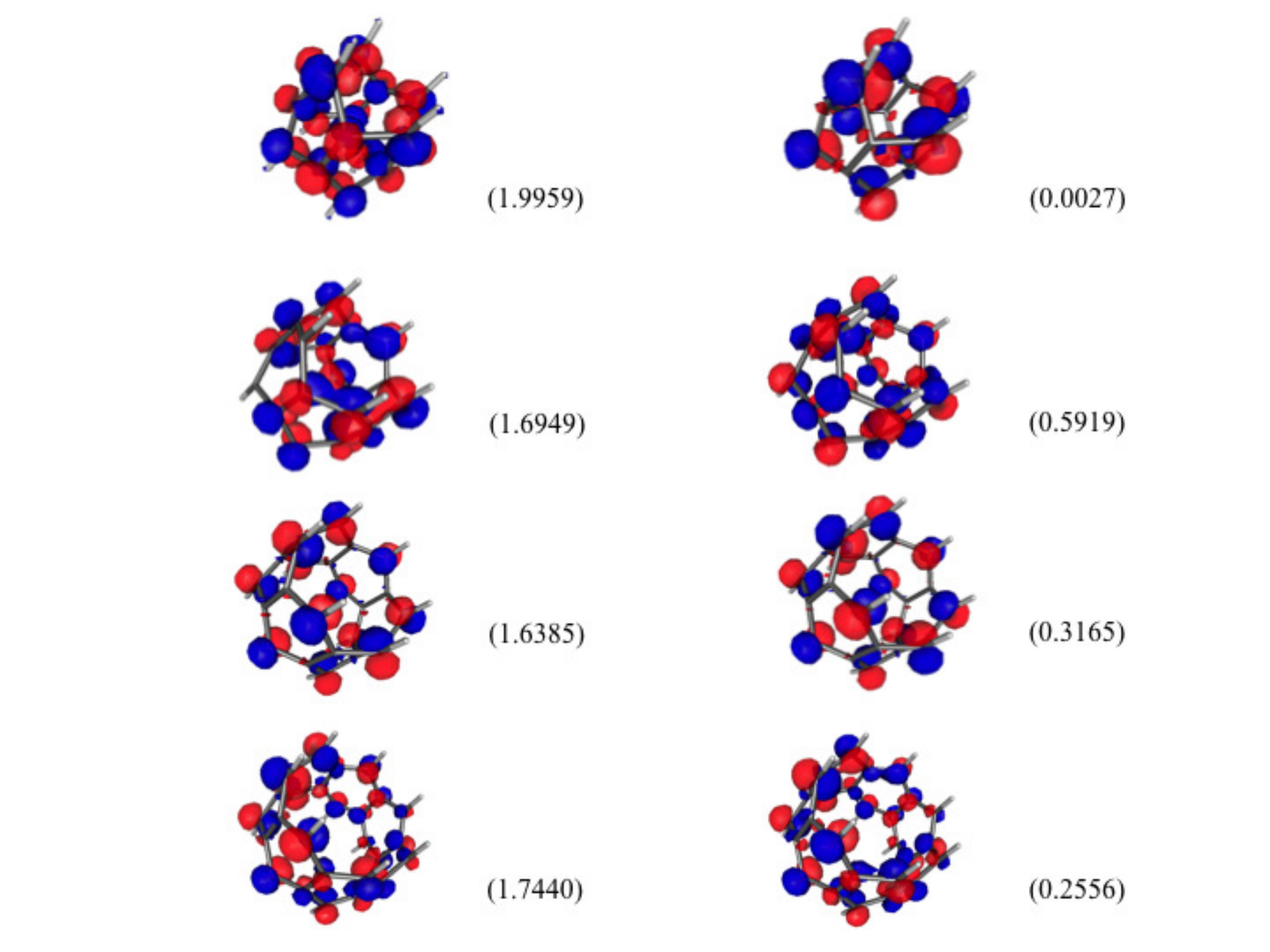} 
\caption{\label{fig:realspace} 
Real-space representation of the HOMOs (left) and LUMOs (right) for the lowest singlet states of 4-cyclacene, 5-cyclacene, 6-cyclacene, and 7-cyclacene, 
calculated using TAO-LDA, at isovalue = 0.01 e/\AA$^3$. The orbital occupation numbers are given in parentheses.} 
\end{figure}

For the lowest singlet states of some representative $n$-cyclacenes ($n$ = 4--7), we explore the real-space representation of active orbitals (e.g., HOMOs and LUMOs), obtained with 
TAO-LDA. Similar to previous findings for $n$-acenes \cite{Acene-DMRG,GNRs-DMRG,GNRs-PHF,z}, the HOMOs and LUMOs of $n$-cyclacenes are mainly localized at the peripheral 
carbon atoms (see \Cref{fig:realspace}).

\section*{Conclusions} 

In conclusion, we have studied the electronic properties of $n$-cyclacenes ($n$ = 4--100), including the ST gaps, vertical ionization potentials, vertical electron affinities, fundamental gaps, 
symmetrized von Neumann entropy, active orbital occupation numbers, and real-space representation of active orbitals, using our newly developed TAO-DFT, a very efficient electronic 
structure method for the study of large systems with strong static correlation effects. To assess the effects of cyclic nature, the electronic properties of $n$-cyclacenes have also been 
compared with those of $n$-acenes. Similar to $n$-acenes, the ground states of $n$-cyclacenes are singlets for all the cases investigated. In contrast to $n$-acenes, the electronic 
properties of $n$-cyclacenes, however, display oscillatory behavior for small $n$ ($n \le 30$) in the approach to the corresponding properties of $n$-acenes with increasing number of 
benzene rings, which to the best of our knowledge have never been addressed in the literature. The oscillatory behavior may be related to the cryptoannulenic effect and structural strain 
of $n$-cyclacene, which have been shown to be important for small $n$, and unimportant for sufficiently large $n$. On the basis of several measures (e.g., the smaller ST gap, the smaller 
$E_{g}$, and the larger $S_{\text{vN}}$), for small $n$, $n$-cyclacene with odd-number benzene rings should possess stronger radical character than that with even-number benzene rings. 
In addition, based on the calculated orbitals and their occupation numbers, the larger $n$-cyclacenes are expected to possess increasing polyradical character in their ground states, where 
the active orbitals are mainly localized at the peripheral carbon atoms. 

Since TAO-DFT is computationally efficient, it appears to be a promising method for studying the electronic properties of large systems with strong static correlation effects. Nevertheless, as 
with all approximate electronic structure methods, a few limitations remain. Relative to the exact full configuration interaction (FCI) method \cite{FCI}, TAO-LDA (with $\theta$ = 7 mhartree) is 
not variationally correct (i.e., overcorrelation can occur), and hence, the orbital occupation numbers from TAO-LDA may not be the same as the NOONs from the FCI method. To assess the 
accuracy of our TAO-LDA results, as the computational cost of the FCI method is prohibitive, the electronic properties of $n$-cyclacenes from relatively affordable {\it ab initio} multi-reference 
methods are called for.

\section*{Acknowledgements} 

This work was supported by the Ministry of Science and Technology of Taiwan (Grant No.\ MOST104-2628-M-002-011-MY3), National Taiwan University (Grant No.\ NTU-CDP-105R7818), 
the Center for Quantum Science and Engineering at NTU (Subproject Nos.:\ NTU-ERP-105R891401 and NTU-ERP-105R891403), and the National Center for Theoretical Sciences of Taiwan. 
We are grateful to the Computer and Information Networking Center at NTU for the partial support of high-performance computing facilities.

\section*{Author Contributions} 
C.-S.W. and P.-Y.L. contributed equally to this work. J.-D.C. conceived and designed the project. C.-S.W. and J.-D.C. performed the calculations. P.-Y.L. and J.-D.C. wrote the paper. 
All authors performed the data analysis.

\section*{Additional Information} 
{\bf Competing financial interests:} The authors declare no competing financial interests.

\newpage 

\newpage 

\newpage 

\newpage 

\newpage 

\newpage 

\newpage 

\newpage 

\newpage 

\newpage 

\end{document}